\documentclass[twocolumn,superscriptaddress,showpacs,preprintnumbers,amsmath,amssymb,prb]{revtex4}

\usepackage{graphicx}
\usepackage{dcolumn}
\usepackage{bm}
\usepackage{amsmath,amssymb}

\begin{document}

\title{Geometrical pumping in spin coupled double quantum dots}
\author{Ryosuke Yoshii}
\affiliation{Yukawa Institute for Theoretical Physics, Kyoto University, Kitashirakawa Oiwake-Cho, Kyoto 606-8502, Japan}
\affiliation{Research and Education Center for Natural Sciences, Keio University, 4-1-1 Hiyoshi, Kanagawa 223-8521, Japan}

\author{Hisao Hayakawa}
\affiliation{Yukawa Institute for Theoretical Physics, Kyoto University, Kitashirakawa Oiwake-Cho, Kyoto 606-8502, Japan}

\date{\today}

\begin{abstract}
We analytically investigate the non-equilibrium pumping for a double quantum dots system on the basis of the quantum master equation (QME), 
where the double quantum dots are connected to two external leads by the spin coupling. Each of leads has two tunable parameters, temperature and chemical potential. 
Using QME formalism, we obtain an analytical expression of the Berry-like phase for eigenstate of the QME in the parameter space. 
We show that the Berry-like curvature is non zero in the whole region in the parameter space. 
We also show that the Berry-like curvature vanishes in the case of 
single quantum dot and the case of isolated two dots. 
\end{abstract}

%05.60.Gg Quantum transport
%72.10.Bg General formulation of transport
%72.15.Qm Scattering mechanisms and Kondo effect
%73.23.-b Electronic transport in mesoscopic systems
%73.63.Kv Quantum dots

\maketitle

\par
\section{Introduction}
Quantum transport phenomenon has been investigated from the early stage of the construction of the quantum theory. 
It still has attracted theoretical and experimental investigations 
to reveal the quantum many-body properties under various non-equilibrium circumstances. 

The recent development of the nanotechnology allows us to make an artificial atom with nanometer scale, 
so called quantum dots, in semiconductor devices. 
There are a number of tunable parameters in the quantum dots, e.g., energy level in a quantum dot, 
bias voltage, and tunnel barrier between the quantum dots and the leads. 
Those tunablities enable us to compare the theory with the experiment for a variety of physical effects. 

One of the most cerebrated development for the quantum transport is the observation of the quantum pump 
in the quantum dot system. 
The original idea of the quantum pumping has been proposed by Thouless.\cite{Thouless} 
When some parameters of the Hamiltonian are adiabatically varied, the eigenstates obtain the Berry phase, which depends on the trajectory in the parameter space and behaves as a vector field.\cite{Berry} 
If the system offers non-zero curvature of the ``vector field" in the parameter space, 
the cyclic modulation of the parameter yields currents originated from the topology of the system. 
The original idea of the quantum pumping for closed system has been extended 
to an open system \cite{Buttiker, Buttiker2, Aleiner, Brouwer, Zhou, Andreev, Cremers, Moskalets, Stefanucci, Brouwer2, Breuer} 
and represented by the geometrical expression. \cite{Makhlin} 
Since then, the various effects on the quantum pumping have been investigated, e.g.\ Rabi oscillation between states of a double quantum dot,\cite{Stafford} interaction of two electrons in a triple-well structure,\cite{Tamborenea} Kondo effect in the Toulouse limit,\cite{Schiller} and so on.   
Experimentally, the quantum pumping has been realized by the transport experiment in the mesoscopic systems.\cite{Kouwenhoven, Pothier, Switkes, Buitelaar, Kaestner, Chorley} 
In those experiments, the quantized dc current has been obtained in the absence of any external bias voltage.

The topological phase also appears in the Master equation.
The parameter of the Liouvillian is adiabatically varied, 
the eigenstates obtain the topological phase similar to the Berry phase. 
This topological phase, so called Berry-Sinitsyn-Nemenman (BSN) phase, has originally offered in the context of the classical master equation for stochastic systems.\cite{Parrondo, Usmani, Astumian, Sinitsyn1, Sinitsyn2, Rahav, Ren, SagawaHayakawa, Breuer} 
In the systems, the cumulant generating function of the pumped current is 
expressed by a Berry-like phase on the eigenstate of the classical master equation.

Recently, the quantum adiabatic pumping on the basis of quantum master equation (QME) has been developed.\cite{Yuge} 
In Ref.\ \onlinecite{Yuge}, they have analyzed the QME for adiabatic modulation of the reservoir parameters, such as temperatures and chemical potentials in leads. 
They have demonstrated that the BSN phase exists in general in the parameter space, and thus, the pumping current can exist in general situations under the adiabatic modulations of reservoir parameters. 
They have applied their method to a double quantum dots system with inter-dot repulsion. 
They obtained the BSN curvature for various interaction strength and have shown that the curvature vanishes in no-interaction limit. 
It is remarkable that the quantum pumping is generated by the modulation of the parameters not for systems but for thermal baths. 
However, their calculation is numerical and explicit parameter dependence is not clear. Moreover, they have not taken into account the spin degrees of freedom in the quantum dots nor the precise intra-dot interaction through localized spins for double dots. 
The BSN curvature in their model has peaks around energy levels of the dots, and that implies the charge fluctuation assists the quantum pumping. 
In this paper, we analyze a more realistic system for double quantum dots. 
We calculate the BSN curvature analytically. 
We have shown that the spin degrees of freedom, or spin fluctuation also play an 
important role in the quantum pumping.

The present paper is organized as follows. 
In Sec.\ II, we introduce the model for the double quantum dot coupled to external leads with spin coupling. 
In Sec.\ III, we make a brief review of calculation for the pumped current in the basis of quantum master equation. 
In Sec.\ IV, we apply the method presented in the Sec.\ III to the double quantum dot system. We derive the analytical expression for the BSN curvature in Sec.\ IV A in leading order of the perturbation. 
In Sec.\ IV B, we calculate the pumped current. 
Finally, in Sec.\ V we summarize and discuss our results.

\section{Model}
In this section we present our model for a system of double quantum dots connected to the source and drain leads. 
In order to clarify the effect of the spin fluctuation on the quantum pumping, 
we focus on the spin localized region in which a number of the electrons in each quantum dot is fixed to one. 
In the region, the system is expressed by the Kondo model where the localized spin 
in the quantum dots couples to the spins in the Fermi see of the leads. 
We assume that the inter-dot coupling is well approximated by the spin coupling of two dots. 

In Appendix A, we demonstrate the absence of the pumping current in the single quantum dot, and the validity of the Kondo model when the temperature is much higher than the characteristic energy, so called Kondo temperature. 

\begin{figure}
\begin{center}
\includegraphics[width=20pc]{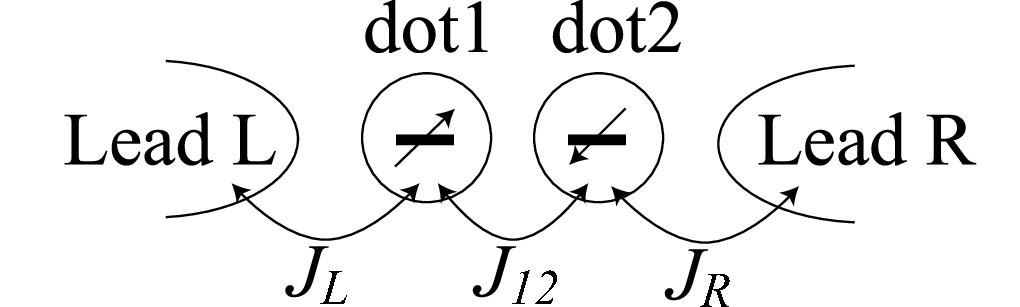}\hspace{2pc}%
\end{center}
\caption{A schematic picture of double quantum dots coupled to external leads}
A double quantum dots with spin $1/2$ in each is connected 
to the external leads. The inter and intra dot interactions are 
considered as the spin coupling. 
We have two control parameters in each lead; 
temperature and chemical potential.
\label{model}
\end{figure}

Our model is depicted in Fig.\ {\ref{model}}. 
The left and right quantum dots are connected with the lead $L$ and lead $R$ by the spin couplings $J_L$ and $J_R$, respectively. 
Here we ignore the energy dependence of $J_L$ and $J_R$ 
by renormalizing the high energy fluctuation into the couplings (see Ref.\ {\onlinecite{Haldane}}). 
Thus Hamiltonian is given by
\begin{eqnarray}
H&=&H_0+H_{\mathrm{K}L}+H_{\mathrm{K}L}+H_{12},\label{H}\\
H_0&=&\sum_{\gamma,k,\sigma}\epsilon_k a_{\gamma,k,\sigma}^\dagger 
a_{\gamma,k,\sigma},\label{H0}\\
H_{\mathrm{K}\alpha}&=&
\sum_{k,k^\prime}J_L\mathbf{S}_1\cdot\mathbf{s}_{L,k,k^\prime}+J_R\mathbf{S}_2\cdot\mathbf{s}_{R,k,k^\prime},\label{H1}\\
H_{12}&=&J_{12}\mathbf{S}_1\cdot\mathbf{S}_2,\label{H12}
\end{eqnarray}
where $a^\dagger_{\gamma,k,\sigma},a_{\gamma,k,\sigma}$ are the 
creation and annihilation operator of the electron with the wave number $k$ and the spin $\sigma$ in the lead $\gamma(=L,R)$, 
$\mathbf{S}_1$ ($\mathbf{S}_2$) is the spin operator in the dot $1$ ($2$), and 
$\mathbf{s}_{\lambda,k,k^\prime}
=\sum_{s,s^\prime}a^\dagger_{\gamma,k,s}
\mathbf{\sigma}_{s,s^\prime}
a_{\gamma,k,s^\prime}$. 
Here we adopt the Hund coupling as the interaction between the quantum dots. 
The derivation of the Hamiltonian (\ref{H})-(\ref{H12}) is presented in Appendix.

We consider an adiabatic modulation of the parameters in the outer region of the double quantum dots. 
Throughout this paper, we treat the temperature in each leads ($T_L$ and $T_R$) and chemical potentials ($\mu_L$ and $\mu_R$) 
as the control parameters. 
It is straightforward to expand the parameter space to include the parameters in inner region of the quantum dots, e.g., energy levels, intra-dot couplings, tunnel potential that corresponds to the modulation of the spin couplings.

%%%%%%%%%%%%%%%%%%%%%%%%%%%%%%%%%%%%%%%%%%%%%%%%%%%%%%%%%%%%%%%%%%%%%
%%%%%%%%%%%%%%%%%%%%%%%%%%%%%%%%%%%%%%%%%%%%%%%%%%%%%%%%%%%%%%%%%%%%%

\section{Method}
In this section, we make a brief review of the method to analyze the 
geometrical pumping in the context of QME.\cite{Yuge}
Here we apply the Keldysh method for calculation of the generating function.\cite{Bagrets, Kindermann} 
We treat the left and right leads as the heat and particle baths. 
After representing the generating function by the path integral, 
we integrate out the Grassmann variables for the leads. 
The general results in the following coincides to that used in Refs.\ \onlinecite{Yuge} and \onlinecite{Esposito}. 
However, that is more tractable and systematic than the method used in Ref.\ \onlinecite{Yuge}.

\begin{figure}
\begin{center}
\includegraphics[width=15pc]{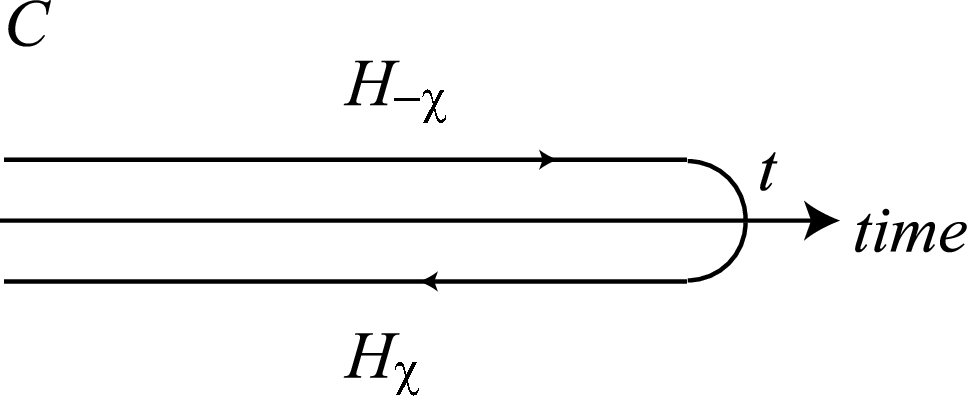}\hspace{2pc}%
\end{center}
\caption{The contour of integral appears in the generating function (\ref{Gf2}).}
\label{contour}
\end{figure}

First, we count the number of the electrons transfer from the left lead to 
the right lead by introducing the counting field. 
The generation function is given by  
\begin{equation}
Z(\chi)=\langle e^{-i\chi N_t}e^{i\chi N_0}\rangle,
\label{Gf}
\end{equation}
where $N_0$ and $N_t$ are the number of the electrons in left leads at initial time $0$ and 
at time $t$, respectively.
For our model, the commutable relation $[H_0,N]=0$ holds. 
Thus Eq.\ (\ref{Gf}) is rewritten as 
\begin{equation}
Z(\chi)=
\mathrm{Tr}
\left[
e^{iH_{\chi} t}
e^{-iH_{-\chi} t} 
\rho_0
\right],
\label{Gf2}
\end{equation}
where $\rho_0$ is the density matrix for initial state and 
\begin{equation}
e^{iH_{\pm\chi} t}=e^{\pm\frac{i}{2}\chi N}
e^{iH t}
e^{\mp\frac{i}{2}\chi N}.
\end{equation}
Equation (\ref{Gf2}) denotes the time evolution depicted in 
Fig.\ \ref{contour}. 
The initial state described by $\rho_0$ evolves by the ``Hamiltonian" 
$H_{-\chi}$ from initial time to $t$ and the state at $t$ evolves backward by $H_{\chi}$. 
%%%%%%%%%%%%%%%%%%%%%%%%%%%%%%%%%%%%%%%%%%%%%%%%%%%%%%%%%%%%%%%%%%%%%
%%%%%%%%%%%%%%%%%%%%%%%%%%%%%%%%%%%%%%%%%%%%%%%%%%%%%%%%%%%%%%%%%%%%%
By using the Keldysh's method of contour integral, we obtain 
\begin{equation}
Z(\chi)=\mathrm{Tr}_{s}T_{c}\int D\xi^\ast D\xi 
\rho_{0s}e^{-i\xi^\ast(G^{-1}+H_{K}^{\chi})\xi},
\end{equation}
where the contour $c$ is represented in Fig.\ \ref{contour}. 
$T_{c}$ is the contour ordering operator defined on $c$.
The bare Green's function is defined by 
\begin{equation}
G=\sigma^z_{\alpha\alpha^\prime}(i\partial_t -H_0)^{-1},
\end{equation}
where $\sigma^z$ represent the $z$ component of the Pauri matrix and $\alpha, \alpha^\prime$ denotes 
the upper contour or lower contour. 
By integrating out the Grassmann field, we obtain  
\begin{equation}
Z(\chi)
=\mathrm{Tr}_{s}T_{c} 
\rho_{0s}e^{\mathrm{Tr}\ln (G^{-1}+H_{K}^{\chi})}.
\end{equation}
Here we can define the density matrix for the system as 
$\rho_s^\chi \equiv T_{c} \rho_{0s}e^{\mathrm{Tr}\ln (G^{-1}+H_{K}^{\chi})}$. 
By using the Markov approximation, the time evolution of $\rho_s^\chi$ becomes local in time and 
thus we obtain the following formal quantum master equation
\begin{equation}
\frac{d\rho_s^\chi}{dt}=\mathcal{K}\rho_s^{\chi}.
\label{QME}
\end{equation}

Next we consider the modulation of the parameters. 
We follow the method described in Ref.\ \onlinecite{Yuge}. 
The density matrix for the system can be expanded as 
\begin{equation}
\rho^\chi(t)=\sum_n c_n(t)e^{\Lambda^\chi_n(t)}\rho^\chi_n(\mathbf{\alpha}_t),
\end{equation}
where $\mathbf{\alpha}_t$ denotes the set of parameters at time $t$, 
$\lambda_n$ and $\rho^\chi_n$ is the eigenvalue and corresponding eigenfunction 
of the $\mathcal{K}$ in Eq.\ (\ref{QME}). 
Here, $\Lambda_n^\chi(t)\equiv \int^t_0 dt^\prime\lambda^\chi_n(\mathbf{\alpha}_t)$ is the dynamical phase corresponds to the house-keeping part of the cumelant generating function. 
Thus we obtain the excess part by subtracting the house-keeping part from the total cumelant generating function. 
The excess part of the cumelant generating function becomes  
\begin{equation}
S_{\mathrm{ex}}(i\chi)=-\int_{C}
\left(
l^\chi_{0,\vec{\alpha}(t)}
,d\rho^\chi_{0,\vec{\alpha}(t)}
\right)+\mathrm{surface\ terms},
\label{Generatingfn}
\end{equation}
where $\rho^\chi_{0,\vec{\alpha}(t)}$ and $l^\chi_{0,\vec{\alpha}(t)}$ 
are the right and left eigenvectors corresponding to the maximum eigenvalue, respectively. 
Here $\vec\alpha(t)$ is the vector which consist of the control parameters. 
Thus we obtain the average number of electrons transfered from the left lead to the system $\Delta N$ by differentiating Eq.\ (\ref{Generatingfn}) with respect to $i\chi$, 
\begin{equation}
\Delta N=-\int_{C}
\left(
l^\prime_{0,\vec{\alpha}(t)}
,d\rho^\chi_{0,\vec{\alpha}(t)}
\right),
\label{excessN}
\end{equation}
where $l^\prime_{0,\vec{\alpha}(t)}=\partial l^\chi_{0,\vec{\alpha}(t)}/\partial(i\chi)|_{\chi=0}$.

\section{Geometrical pumping}
In this section, we apply the method described in the previous section to the model depicted in Fig.\ 1. 
In the first subsection, we derive the QME for our system. 
We apply the Markovian approximation and perturbation for spin coupling, 
we obtain the QME with counting field. 
In the second subsection, we calculate the BSN phase and curvature for our system. 
We also estimate the pumping current for cyclic modulation. 
\subsection{Quantum Master Equation}
We introduce the counting fields $\chi_{\gamma\sigma}$ to count the number of the electrons $N_{\gamma\sigma}$ with spin $\sigma$ in lead $\gamma$. 
The generating function becomes 
\begin{equation}
Z(\chi)
=\mathrm{Tr}_{s}T_{c} 
\rho_{0s}e^{\mathrm{Tr}\ln (G^{-1}+H_{L,K}^{\chi}+H_{R,K}^{\chi}+H_{12})},
\label{CGF}
\end{equation}
with
\begin{equation}
H_{\gamma,\mathrm{K}}^\chi=
J_\gamma\mathbf{s}^\gamma_{\sigma,\sigma^\prime}\cdot
\mathbf{S}^{\delta(\gamma)\alpha}_{\omega\omega^\prime}
\sigma_{\alpha\alpha^\prime}^z
e^{-i(\chi_{\gamma\sigma}-\chi_{\gamma\sigma^\prime})
\sigma^z_{\alpha\alpha^\prime}/2},
\end{equation}
where $\alpha=+$ or $-$ represents the Keldysh branch and 
$\mathbf{S}^{\delta(\gamma)\alpha}_{\omega\omega^\prime}
=\int dt e^{-i(\omega-\omega^\prime)t}\mathbf{S}^{\delta(\gamma)\alpha}_a(t)$. 
Here we define $\delta(L)=1, \delta(R)=2$.
The Hund coupling term $H_{12}$ is commutable with the number operator for leads. 
We focus on the case of $\pi\nu J_\gamma \ll 1$ and $J_{LR}/J_\gamma\ll 1$. 
Thus we make the perturbation of $J_{\gamma}$ and $J_{12}$ for Eq.\ (\ref{CGF}). 
We can show that the off diagonal part of the density matrix is irrelevant in the steady state. 
In the case of adiabatic modulation we have assumed, the contribution of these terms is negligible. 
Thus the QME is closed in the diagonal part of the density matrix, $\rho^\chi_{\sigma\sigma^\prime}$, 
where $\sigma$ and $\sigma^\prime$ stand for the spin in the left and right dots, respectively. 
By using the Markovian approximation, 
the QME for the diagonal part of $\rho^\chi_{s}$ 
becomes 
\begin{equation}
\partial_t \rho^\chi_s=R^\chi \rho^\chi_s,
\label{QME2}
\end{equation}
\begin{equation}
R^\chi=\left(
\begin{array}{cccc}
a&b_R^+&b_L^+&0\\
b_R^-&a&-J_{LR}&b_L^+\\
b_L^-&-J_{LR}&a&b_R^+\\
0&b_L^-&b_R^-&a
\end{array}
\right),
\label{QM1}
\end{equation}
in the leading order of the perturbation, 
where, $\rho^\chi_s=(\rho_{\uparrow\uparrow},\rho_{\uparrow\downarrow},\rho_{\downarrow\uparrow},
\rho_{\downarrow\downarrow})^{t}$ and 
\begin{eqnarray}
a&=&\frac{\pi^2\nu^2}{4}\sum_{\gamma,\gamma^\prime}\int\frac{d\epsilon}{2\pi}
J_{\gamma}J_{\gamma^\prime}\left[1-f_{\gamma}(\epsilon)\right]f_{\gamma^\prime}(\epsilon)\nonumber\\
&&\times
\left[\left(e^{-i(\chi_{\gamma\uparrow}-\chi_{\gamma^\prime\uparrow})}
+e^{-i(\chi_{\gamma\downarrow}-\chi_{\gamma\prime\downarrow})}\right)-6\right],\label{diag}\\
b_\gamma ^+
&=&
(\pi\nu J_{\gamma})^2
\int\frac{d\epsilon}{2\pi}
\left[1-f_{\gamma\uparrow}(\epsilon)\right]f_{\gamma\downarrow}(\epsilon)
e^{-i(\chi_{\gamma\uparrow}-\chi_{\gamma\downarrow})},
\label{offdiag}
\\
b_\gamma^-
&=&
(\pi\nu J_{\gamma})^2
\int\frac{d\epsilon}{2\pi}
\left[1-f_{\gamma\downarrow}(\epsilon)\right]f_{\gamma\uparrow}(\epsilon)
e^{-i(\chi_{\gamma\downarrow}-\chi_{\gamma\uparrow})},
\label{offdiag2}
\end{eqnarray}
with Fermi distribution function $f_\gamma(\epsilon)=\left[1-e^{-\beta_\gamma(\epsilon-\mu_\gamma)}\right]^{-1}$ in the lead $\gamma$. 
Here $\beta_\gamma=T_\gamma^{-1}$ and $\mu_\gamma$ are the inverse temperature and the chemical potential in the lead $\gamma$, respectively. 
As we mentioned above, here we consider the parameters to be $\vec{\alpha}(t)=(\mu_L, T_L, \mu_R, T_R)$. 

\subsection{BSN curvature and pump current}
\begin{figure}
\begin{center}
\includegraphics[width=10pc]{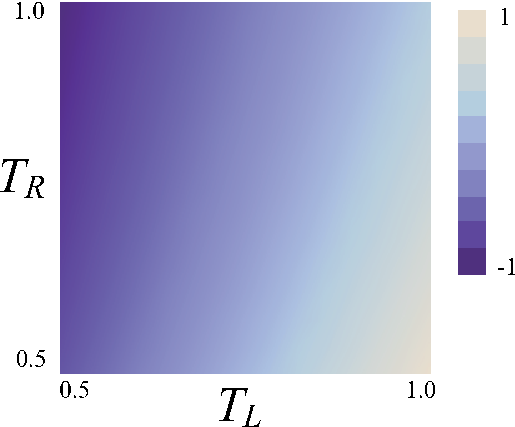}\hspace{2pc}%
\end{center}
\caption{The density plot of curvature $F_{\beta_L\beta_R}/\alpha_{12}$ in Eq.\ (\ref{BSNC}) in a case of $J_L=J_R$.  
The curvature is nonzero in the temperature region where the temperature is much higher than the Kondo temperature. }
\label{figcurvature}
\end{figure}

It is easy to show that the QME (\ref{QM1}) has four eigenvalues $\lambda_a$ ($a=0,1,2,3$), 
\begin{eqnarray}
\lambda_0=\frac{1}{2}\left[2a+(b_L^-+b_L^+)+\sqrt{(b_L^--b_L^+)^2+4b_R^-b_R^+}\right],\\
\lambda_1=\frac{1}{2}\left[2a-(b_L^-+b_L^+)+\sqrt{(b_L^--b_L^+)^2+4b_R^-b_R^+}\right],\\
\lambda_2=\frac{1}{2}\left[2a+(b_L^-+b_L^+)-\sqrt{(b_L^--b_L^+)^2+4b_R^-b_R^+}\right],\\
\lambda_3=\frac{1}{2}\left[2a-(b_L^-+b_L^+)-\sqrt{(b_L^--b_L^+)^2+4b_R^-b_R^+}\right].
\end{eqnarray} 
It is obvious that the eigenvalue $\lambda_0$ has the largest real part. 
When $\chi=0$, the right eigenstates corresponding to $\lambda_0$ becomes, 
\begin{equation}
\rho_0=\frac{1}{4}\left(1+A, 1-A, 1-A, 1+A\right)^{t}, 
\label{RES}
\end{equation}
where, $A={2J_{LR}}/({b_L+b_R})$ is dimensionless parameter with 
$b_L=b_L^\pm|_{\chi=0}, b_R=b_R^\pm|_{\chi=0}$ and $t$ represents the transverse. 
The quantities $b_L$ and $b_R$ can be calculated by Eq.\ (\ref{offdiag}) 
as
\begin{equation}
b_\gamma=(\pi\nu J_\gamma)^2\frac{T_\gamma}{2\pi}, 
\label{bbeta}
\end{equation}
The derivative of left eigenvalue with respect to $i\chi_{L\uparrow}$ 
at $\chi=0$ which enters in Eq.\ (\ref{excessN}) becomes,
\begin{equation}
l^\prime_0=\left(-\frac{b_R}{b_L}, 0, \frac{b_R}{b_L}, 0\right).
\label{LES}
\end{equation}
The BSN curvature for our model formally satisfies
\begin{equation}
\sum_{x,y} F_{x,y}dx \wedge dy 
=dl^\prime_0\wedge d\rho_0,
\label{curvature}
\end{equation}
where $x, y=\mu_L, \mu_R, \beta_L, \beta_R,$ and the total derivative in the right hand side is taken in the parameter space and $\wedge$ is the edge product. 
Substituting Eqs.\ (\ref{RES})-(\ref{LES}) into Eq.\ (\ref{curvature}), the BSN curvature becomes 
\begin{equation}
F_{\beta_L\beta_R}=\frac{\alpha_{12} T_L^2}{T_L+\beta_{LR}^2T_R},
\label{BSNC}
\end{equation} 
where $\alpha_{12}={J_{12}}/{(\pi\nu J_R)^2}$ and $\beta_{LR}=J_{R}/J_{L}$. 
This expression of the curvature is proportional to $T_L$ at high $T_L$. 
It is notable that the curvature always survives thanks to the spin coupling. We should note, however, that our model is valid for spin localized region i.e.\ for $T_L, T_R \ll U$, where $U$ is the intra-dot interaction. 
Thus we expect that the Eq.\ (\ref{BSNC}) saturates to finite value. 
Our model in which the two Kondo impurities are coupled with spin coupling requires 
the condition $J_{12}\ll J_{L}, J_{R}$. 
In Fig.\ \ref{figcurvature}, we show the curvature Eq.\ (\ref{BSNC}). 
The curvature is non-zero in the whole parameter region where the temperature is much higher than the Kondo temperature.

%%%%%%%%%%%%%%%%%%%%%%%%%%%%%%%%%%%%%%%%%%%%%%%%%%%%%%%%%%%%%%%%%%%%%
%%%%%%%%%%%%%%%%%%%%%%%%%%%%%%%%%%%%%%%%%%%%%%%%%%%%%%%%%%%%%%%%%%%%%

In our system, the BSN curvature vanishes in the chemical potential plane. 
This result seems completely different from the results in Ref.\ \onlinecite{Yuge}, which offers non-zero curvature in chemical potential plane. 
However, the result we obtain above is consistent with the previous results 
by following reason. 
The BSN curvature in Ref.\ \onlinecite{Yuge} has peaks around energy levels $\epsilon_1, \epsilon_2$ in quantum dots and $\epsilon_1+U, \epsilon_2+U$. 
The result clearly shows that the pumping 
is assisted by the charge fluctuation. 
In this paper, however, we restrict our model to the spin localized region by 
suppressing the charge fluctuation in construction of Kondo model. 
Thus our model does not allow us to describe the quantum pumping assisted by charge fluctuation which is dominant in chemical potential modulation. 
In our case, the spin fluctuation assists the quantum pumping. 

\subsection{Pumping current}
In this subsection, we explicitly calculate the pumped electrons in the left lead, 
$\Delta N_{L\uparrow}$ generated from $t=0$ to $t_0$. 
In order to calculate $\Delta N_{L\uparrow}$ we set $\chi_{L\uparrow}\equiv \chi$ and $\chi_{L\downarrow}=0$, $\chi_{R\sigma}=0$ as in the previous section. 
Thus we obtain $l^\prime_0=\left(-{b_R}/{b_L}, 0, {b_R}/{b_L}, 0\right)$.
Substituting the above expression of $l^\prime_0$ into Eqs.\ (\ref{RES})-(\ref{bbeta}) 
into Eq.\ (\ref{excessN}), we obtain 
\begin{equation}
\Delta N_{L\uparrow}=J_{LR}\int \frac{b_R}{b_L}d(b_L+b_R)^{-1}.
\label{pumpN}
\end{equation} 
%Here we mention, if we calculate the charge pump by setting 
%$\chi_{L\uparrow}=\chi_{L\downarrow}$, we obtain $\Delta N_{L}\equiv \Delta(N_L+N_R)=0$. 

\begin{figure}
\begin{center}
\includegraphics[width=8pc]{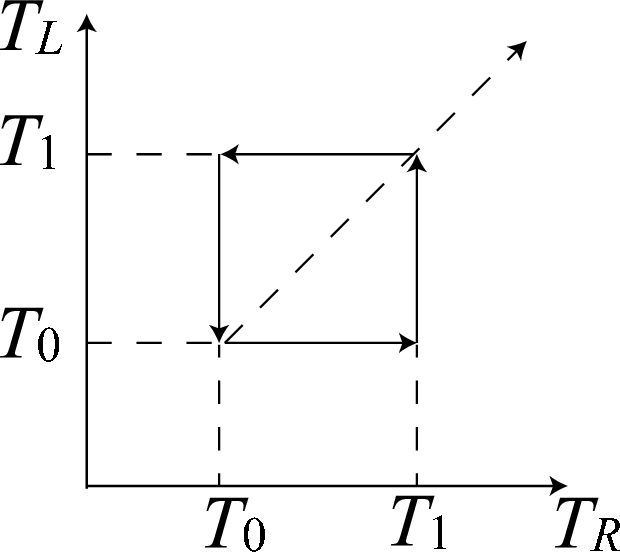}%
\includegraphics[width=12pc]{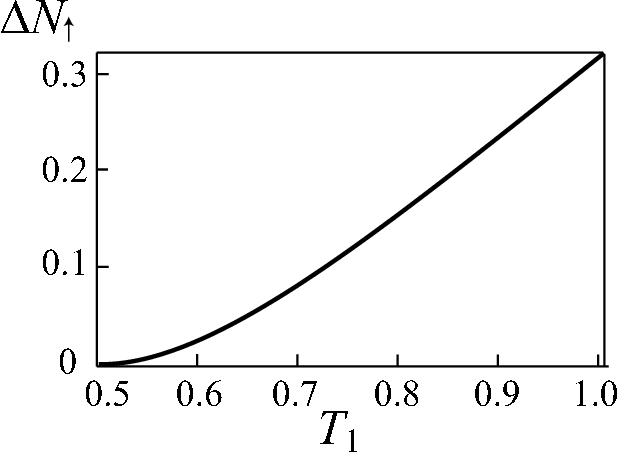}%
\end{center}
\caption{The excess current for square modulation in temperature plane of the parameter space. 
The left figure shows the trajectory in the temperature plane. 
We start from $(T_L, T_R)=(T_0, T_0)$ and increase $T_R$ from $T_0$ to $T_1$. 
Next we change $T_L$ as $T_0\rightarrow T_1$. 
Then we modulate the temperature from $(T_1, T_1)$ to $(T_0, T_1)$ by lowering $T_L$ and 
go back to original point by lowering $T_R$.
Here we set $J_L=J_R$, $T_0=0.5$ and $\alpha_{12}/2T_0=1$. 
The right figure shows the $T_1$ dependence of the excess current. 
We show that the larger area enclosed by the trajectory achieves larger creation of the excess current.}
\label{figexcess}
\end{figure}

As we show in the previous subsection, the BSN curvature in the parameter space is non-zero. 
Thus $\Delta N_{L\uparrow}$ obtained by (\ref{pumpN}) depends on the trajectory in the parameter space. 
As an example, we calculate the excess current for modulation depicted in left figure of Fig.\ {\ref{figexcess}}. 
We change the temperature in left and right leads from $(T_0, T_0)$ to $(T_1, T_1)$ 
by the square trajectory in Fig.\ {\ref{figexcess}}. 
In the right figure of Fig.\ {\ref{figexcess}}, we show the $T_1$ dependence of the excess current 
achieved by the one cycle. 
The results shows that we can obtain the large excess current by 
raising $T_1$. 
Here we again mention that the range of the temperatures $T_0$ and $T_1$ 
are restricted in the region much larger than $T_{\mathrm{K}}$ and much less than 
$U$.
%%%%%%%%%%%%%%%%%%%%%%%%%%%%%%%%%%%%%%%%%%%%%%%%%%%%%%%%%%%%%%%%%%%%%
%%%%%%%%%%%%%%%%%%%%%%%%%%%%%%%%%%%%%%%%%%%%%%%%%%%%%%%%%%%%%%%%%%%%%

\section{Conclusion}
We have derived the geometrical pumping for a system of double quantum dots connected with heat baths by external leads based on the QME approach. 
In the spin localized region, the spin fluctuation is dominant for the quantum pumping 
and that is complementary to the result for a spin less model in Ref.\ \onlinecite{Yuge} where the 
charge fluctuation is dominant. 
We have also obtained the analytical expression of BSN curvature in the parameter space for our model. 
We have shown that the geometrical pumping is not realized by the 
single dot nor the isolated two dots (double dot without inter dot repulsion) by the modulation of temperature or chemical potential, where 
each dot is restricted to the spin localized region and has spin degeneracy. According to our analysis, if we destroy the spin degeneracy by applying the magnetic field or connected to the ferromagnetic leads, the spin pump can be yielded. 
In contrast to that, we obtain the pumping induced by the inter dot repulsion. 
In this case, we obtain non zero pumping by modulation of the temperature in the external leads.

In this paper we restrict our analysis to the quantum dots in the spin localized region (weak coupling region). 
However, the QME formalism is applicable to the whole region of the coupling strength. 
In the strong coupling region, the Kondo singlet state is formed around the Fermi surface, that is not captured by the spinless model in Ref.\ \onlinecite{Yuge} nor the weak coupling model in our paper. 
The Kondo singlet state is well known to bring a drastic change to transport phenomena at the temperatures lower than the Kondo temperature. 
Concerning to that, the spin pumping in the Kondo regime for Toulouse limit has been proposed recently.\cite{Schiller} 
It is a future issue to reveal the effect of the many body correlation on the 
quantum pump. 
The investigation of the non-adiabatic correction and non-Markovian effect are also the future issues. 
In the case of boson transport, the non-adiabatic effect and non-Markovian effect have been studied and 
the generalization to the case of fermion transport is straightforward.\cite{Uchiyama}

\section*{Acknowledgment}
The authors acknowledge T.\ Sagawa, K.\ Nakata, and T.\ Yuge for their helpful advices and K.\ Watanabe for valuable discussion. 

\appendix

\section{Case of single dot}
\begin{figure}
\begin{center}
\includegraphics[width=20pc]{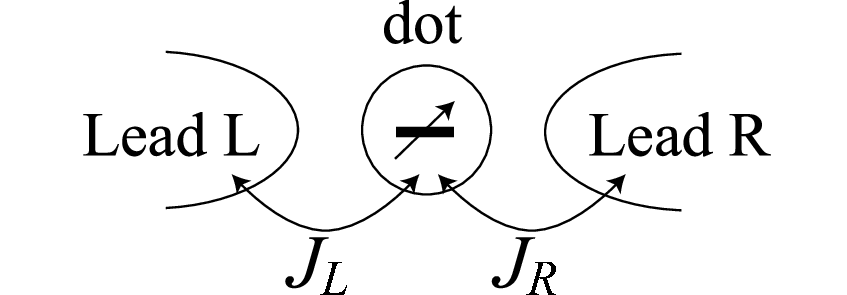}\hspace{2pc}%
\end{center}
\caption{Model for the single dot connected to the leads.}
\label{singledot}
\end{figure}

We devote this appendix to show that the geometrical pumping is 
prohibited for the single quantum dot system. 
Our model analyzed in this appendix is depicted in Fig.\ \ref{singledot}. 

%%%%%%%%%%%%%%%%%%%%%%%%%%%%%%%%%%%%%%%%%%%%%%%%%%%%%%%%%%%%%%%%%%%%%
%%%%%%%%%%%%%%%%%%%%%%%%%%%%%%%%%%%%%%%%%%%%%%%%%%%%%%%%%%%%%%%%%%%%%

The Hamiltonian is given by
\begin{eqnarray}
H&=&H_0+H_{\mathrm{int}},\\
H_0&=&\sum_{\alpha,k,\sigma}\epsilon_k a_{\alpha,k,\sigma} a_{\alpha,k,\sigma}
    +\sum_{\sigma}\epsilon_0 d^\dagger_{\sigma} d_\sigma
    +Un_{\uparrow}n_{\downarrow},\\
H_{\mathrm{T}}&=&\sum_{\alpha,\sigma}V_{\alpha} d^\dagger_{\sigma}a_L+\mathrm{h.c.}\ ,
\end{eqnarray}
where, $a_{\alpha, k,\sigma}^\dagger$ and $a_{\alpha, k,\sigma}$ are the creation and the annihilation operator 
for the electron in the leads $\alpha(=L,R)$ with wave number $k$, energy $\epsilon_k$, and spin $\sigma$.
$d^\dagger_\sigma,d_\sigma$ is those in the quantum dot and  
$n_{\sigma}=d^\dagger_{\sigma}d_{\sigma}$. 
$U$ is the electron-electron interaction in the quantum dot.

By using the Schriefer-Wolff transformation\cite{Schriefer-Wolff}, 
the above model reduces to the following model:
\begin{eqnarray}
H&=&H_0+H_{\mathrm{K}},\label{HKondo}\\
H_0&=&\sum_{k,\sigma}\epsilon_k a_{k,\sigma} a_{k,\sigma},\\
H_{\mathrm{K}}&=&\sum_{k,k^\prime}J\mathbf{S}\cdot\mathbf{s}_{k,k^\prime},
\end{eqnarray}
where, $a_{k,\sigma}=V_{L}a_{L,k,\sigma}+V_{R}a_{R,k,\sigma}/V$, $ V=\sqrt{|V_L|^2+|V_R|^2}$ and 
$\mathbf{S}$ is the spin operator in the quantum dot. 
We denote the spin operator for the conduction electrons as 
$\mathbf{s}_{k,k^\prime}=\sum_{\sigma,\sigma^\prime}
a_{k,\sigma}^\dagger a_{k^\prime,\sigma^\prime}
\mathbf{\sigma}_{\sigma,\sigma^\prime}/2$.
The coupling constant $J$ is presented by the density of states in leads $\nu$ as $J=2\pi \nu V^2/(\epsilon_0+U)$. 
We assume the density of states to be constant (wide-band limit).
Here we focus on the electron-hole symmetry ($|\epsilon_0|=\epsilon_0+U$) for simplicity. 
\cite{note1}

First, we renormalize the spin fluctuation from the 
high energy region. 
We apply the poor man's scaling method \cite{Anderson, Haldane} to 
the Kondo model Eq.\ (\ref{HKondo}). 
This yields the effective coupling 
\begin{equation}
\tilde J=\frac{1}{2\nu \ln \mathrm{max}\{eV,T\}/T_{\mathrm{K}}},
\label{tilJ}
\end{equation}
where $V$ is the bias voltage between the left lead and the right lead.
The Kondo temperature $T_{\mathrm{K}}$ is defined by 
\begin{equation}
T_{\mathrm{K}}=D_0\exp\left(\frac{1}{2\nu J}\right).
\end{equation}
In this paper, we focus on the spin localized region (or the weak coupling region) $\nu \tilde J\ll 1$. 
In the region, the conduction electrons are scattered by the local spin in the dot sequentially. 

In this case, 
the QME for the diagonal parts of $\rho^\chi_{s}$ 
becomes 
\begin{equation}
\partial_t \rho^\chi_s=R^\chi \rho^\chi_s,
\end{equation}
\begin{equation}
R^\chi=\left(
\begin{array}{cc}
a&(\pi\nu J)^2 b^+\\
(\pi\nu J)^2 b^-&a\\
\end{array}
\right),
\end{equation}
in the leading order of the perturbation in $J$ by using the Markov approximation. 
Here $a$ is the given by Eq.\ (\ref{diag}) with $J_L=J_R=J$ and 
\begin{eqnarray}
b^+&=&\sum_{\gamma,\gamma^\prime}\int\frac{d\epsilon}{2\pi}
\left[1-f_{\gamma}(\epsilon)\right]f_{\gamma^\prime}(\epsilon)e^{-i(\chi_{\gamma\uparrow}-\chi_{\gamma^\prime\downarrow})},\\
b^-&=&\sum_{\gamma,\gamma^\prime}\int\frac{d\epsilon}{2\pi}
\left[1-f_{\gamma}(\epsilon)\right]f_{\gamma^\prime}(\epsilon)e^{-i(\chi_{\gamma\downarrow}-\chi_{\gamma^\prime\uparrow})}.
\end{eqnarray}
Thus we obtain the right and left eigenstates corresponding to the eigenvalue 
with largest real part as
\begin{eqnarray}
\rho^\chi_{\vec{\alpha}(t)}
&=&\frac{1}{2}
\left(
\begin{array}{c}
\sqrt{b^+/b^-}\\
1
\end{array}
\right),\label{lev}\\
l^\chi_{\vec{\alpha}(t)}
&=&
\left(
\begin{array}{cc}
\sqrt{b^-/b^+}&
1
\end{array}
\right).\label{rev}
\end{eqnarray}
Taking the derivative with parameters for $\rho^\chi_{\vec\alpha(t)}$ yields 
\begin{eqnarray}
d\rho^\chi_{\vec{\alpha}(t)}
&=&\frac{1}{\sqrt{2}}d\vec{\alpha}\cdot
\left(
\begin{array}{c}
\frac{d}{d\vec{\alpha}}(\sqrt{b^+/
b^-})\\
0
\end{array}
\right).
\end{eqnarray}
By setting $\chi=0$ and using the property $b^+|_{\chi=0}=b^-|_{\chi=0}$, 
we obtain 
\begin{eqnarray}
\left.d\rho^\chi_{\vec{\alpha}(t)}\right|_{\chi=0}
&=&0.
\end{eqnarray}
Thus the curvature vanishes in the parameter space. 


\begin{thebibliography}{}
\bibitem{note1}When the electron-hole symmetry is broken, the additional 
potential scattering term $\sum_{k,k^\prime} K a^\dagger_{k,\sigma}a_{k^\prime,\sigma}$ 
appears in the Kondo Hamiltonian. 
However this term is irrelevant for the renormalization, whereas the spin exchange term is relevant. 
Thus we can neglect the potential scattering term.
\end{thebibliography}
\end{document}